\documentclass[twocolumn,aps,prb,superscriptaddress,amsmath,amssymb,showpacs,floatfix,preprintnumbers]{revtex4-2}
\usepackage[latin9,utf8]{inputenc}
\usepackage{float}
\usepackage{textcomp}
\usepackage{amsmath,amssymb,amscd,hyperref}
\usepackage{graphicx}
\makeatletter
\usepackage{url}
\usepackage{grffile}
\usepackage{bbold}
\usepackage{comment}
\usepackage{dirtytalk}
\usepackage{chngcntr}
\usepackage{tabularx}
\usepackage{multirow}
\usepackage{soul}
\usepackage{cleveref}
\usepackage{braket}
\usepackage{makecell}  
\usepackage{amsmath}   
\usepackage{graphicx}  
\DeclareUnicodeCharacter{2212}{-}
\DeclareUnicodeCharacter{0301}{\hspace{-1ex}\'{ }}

\usepackage{dcolumn}
\usepackage{color}
\DeclareGraphicsExtensions{.png .jpg .pdf}
\usepackage{hyperref}
\hypersetup{
     colorlinks = true,
     linkcolor = blue,
     anchorcolor = blue,
     citecolor = blue,  
     filecolor = blue,
     urlcolor = blue
     }
\usepackage{braket}
\usepackage{physics}
\usepackage{array}
\usepackage{booktabs}
\usepackage{soul}

\makeatother
\begin{document}   

%\preprint{AIP/123-QED}
%\title{Probing the Rashba effect in monolayer transition metal dichalcogenides by plasmons}

\title{Plasmonic detection of Rashba spin-orbit coupling in monolayer transition metal dichalcogenides}

\author{Y. Li}
\affiliation{School of Physics and Astronomy, Yunnan University, Kunming 650091, People’s Republic of China}
\address{Department of Physics and NANOlight Center of Excellence, University of Antwerp, Groenenborgerlaan 171, 2020 Antwerp, Belgium}

\author{Z. H. Tao }
\email{zehua.tao@uantwerpen.be}
\address{Department of Physics and NANOlight Center of Excellence, University of Antwerp, Groenenborgerlaan 171, 2020 Antwerp, Belgium}

\author{Y. M. Xiao}
\email{yiming.xiao@ynu.edu.cn}
\affiliation{School of Physics and Astronomy, Yunnan University, Kunming 650091, People’s Republic of China}

\author{W. Xu}
\email{wenxu\_issp@aliyun.com}
\affiliation{School of Physics and Astronomy, Yunnan University, Kunming 650091, People’s Republic of China}
\affiliation{Key Laboratory of Materials Physics, Institute of Solid State
Physics, HFIPS, Chinese Academy of Sciences, Hefei 230031,  P. R. China}

\author{Q. N. Li}
\affiliation{School of Physics and Astronomy, Yunnan University, Kunming 650091, People’s Republic of China}

\author{F. M. Peeters}
 \address{Department of Physics and NANOlight Center of Excellence, University of Antwerp, Groenenborgerlaan 171, 2020 Antwerp, Belgium}
\affiliation{Departamento de F\'{i}sica, Universidade Federal do Cear\'{a}, Caixa Postal 6030, Campus do Pici, 60455-900 Fortaleza, Cear\'{a}, Brazil}

\author{D. Neilson}
 \address{Department of Physics and NANOlight Center of Excellence, University of Antwerp, Groenenborgerlaan 171, 2020 Antwerp, Belgium}

\author{M. V. Milo\v{s}evi\'{c}}
\email{milorad.milosevic@uantwerpen.be}
\address{Department of Physics and NANOlight Center of Excellence, University of Antwerp, Groenenborgerlaan 171, 2020 Antwerp, Belgium}

\begin{abstract}
Rashba spin-orbit coupling (RSOC) induces strong momentum-dependent spin splitting and plays a crucial role in fields such as spintronics and topological photonics. We here theoretically investigate the collective excitations in monolayer transition metal dichalcogenides (ML-TMDs) hosting RSOC, and conceive an approach to precisely quantify the strength of RSOC using plasmons. We determine the electron energy loss function (EELF) and plasmon dispersions for $n$-type ML-TMDs from the dynamic dielectric function in the framework of the standard random phase approximation. In this system, both optical and acoustic plasmon modes are observed in the EELF and plasmon dispersions. Moreover, the plasmonic and spectral properties are tunable by electron density and dependent on RSOC. Crucially, we identify a minimum energy gap between the two plasmon modes to serve as a direct spectral signature of the RSOC strength. These results establish plasmons as a noninvasive, precise, and broadly tunable technique for determining RSOC in TMD van der Waals heterostructures and devices.
\end{abstract}

\maketitle
{\it Introduction.}
The Spin-orbit coupling (SOC) effect offers a paradigmatic platform for controlling the electron spin. Rashba spin-orbit coupling (RSOC) emerges in inversion asymmetric materials~\cite{shanavas2014theoretical,manchon2015new}, and introduces a momentum-dependent effective magnetic field experienced by moving electrons. This leads to spin-momentum locking~\cite{manchon2015new}, which is a fundamental coupling between the direction of an electron's momentum and its spin orientation. Such a locking mechanism is not only essential for spintronics applications ~\cite{manchon2015new,koo2020rashba} but also has profound implications for topological photonics~\cite{wang2025inherent,shi2025efficient} and enhanced superconductivity~\cite{patterson2025superconductivity}.

Two-dimensional (2D) materials readily enable controllable RSOC effects, as their inversion symmetry can be broken either intrinsically through structural design~\cite{feng2025giant} or externally via applied electric field~\cite{xu2012electric,ho2015gate,patel2022electric}. Monolayer transition metal dichalcogenides (ML-TMDs) are of special interest in that respect. They possess a hexagonal lattice structure that naturally breaks inversion symmetry, and the presence of heavy transition metal atoms results in strong intrinsic SOC (iSOC)~\cite{zhu2011giant}. The iSOC leads to unique effects such as spin-valley coupling~\cite{xiao2012,zeng2013optical} and light-valley coupling~\cite{mak2018light}. Moreover, the application of an electric field in ML-TMDs can further enhance inversion asymmetry, enabling dynamic modulation of the RSOC strength through gate voltage~\cite{patel2022electric}. Therefore, ML-TMDs provide an ideal platform for further investigating and controlling the RSOC and its effects. %The exploration of RSOC in these materials promises to deepen our understanding of spintronics and establish a theoretical foundation for spintronic devices.

Previous theoretical studies have predicted that RSOC influences optical conductivity \cite{maytorena2006spin,Xiao16} and plasmon \cite{badalyan2009anisotropic} response but did not provide a quantitative method to determine the RSOC strength. Existing techniques for evaluating SOC such as electrical transport measurements~\cite{ast2007local,abrao2024probing}, angle-resolved photoemission spectroscopy (ARPES), and spin-resolved ARPES (spin-ARPES)~\cite{feng2025giant} have several limitations. For instance, electrical techniques cannot directly measure SOC; instead, it is typically extracted by measuring spin currents. In addition, these techniques require deposition of metal leads, which is often at odds with fragility of 2D materials. ARPES and spin-ARPES often involve the transfer of monolayers onto conductive substrates, risking strain and contamination. Therefore, these methods are not suited for large-scale, noninvasive, but precise characterization in an industrial context.

To solve this unfavorable issue, we propose an optical, contact-free, and nondestructive method to probe RSOC strength in ML-TMDs, using plasmons. Plasmons strongly couple with light and are highly sensitive to the electronic and structural properties of materials ~\cite{papaj2023probing,lavor2020probing,tao2021tailoring,tao2023ultrastrong,tao2024chiral}. Our approach employs the unique plasmonic response caused by RSOC-induced spin hybridization and terahertz intersubband electronic transitions, offering a reliable and scalable way to detect RSOC strength, without harming the sample. This technique is therefore promising for large-scale probing of RSOC strength in practical environment and applications.

{\it Theoretical approach. }
We consider a ML-TMD laid on a substrate in the $x$-$y$ plane. The effective Hamiltonian for an electron near the $K$ ($K'$) valley, including the RSOC, can then be written as~\cite{Qi15, Xiao16, liu2024}
\begin{align}\label{hamil}
\hat{H}^\zeta=\hat{H}_0+\hat{H}_{iSOC}+\hat{H}_R,
\end{align}
with
\begin{align}
&\hat{H}_0=[at(\zeta k_x \hat{\sigma}_x+k_y \hat{\sigma}_y)+\Delta \hat{\sigma}_z/2]\otimes\hat{s}_0, \tag{1a}\\
&\hat{H}_{iSOC}=\zeta (\lambda_c \hat{\sigma}_++\lambda_v \hat{\sigma}_-)\otimes \hat{s}_z, \tag{1b}\\
&\hat{H}_{R}=\lambda_R(\zeta \hat{\sigma}_x \otimes \hat{s}_y-\hat{\sigma}_y \otimes \hat{s}_x), \tag{1c}
\end{align}
% as a $4 \times 4$ matrix
% \begin{align}
% \hat{H}^\zeta =
% {\small
% \begin{pmatrix}
% \Delta_{c \zeta}^{+} & 0 & atk^- & -i(\zeta-1)\lambda_R \\
% 0 & \Delta_{c \zeta}^{-} & i(\zeta+1)\lambda_R & atk^- \\
% atk^+ & -i(\zeta+1)\lambda_R & -\Delta_{v \zeta}^{-} & 0 \\
% i(\zeta-1)\lambda_R & atk^+ & 0 & -\Delta_{v \zeta}^{+}
% \end{pmatrix}
% },
% \end{align}
%
where $\mathbf{k} = (k_x, k_y)$ is the in-plane wave vector, $k^\pm = \zeta k_x \pm i k_y$, and $\zeta = \pm $ indexes the $K$ and $K'$ valleys. $\hat{\sigma}_i$ represents the Pauli matrices
of the sublattice pseudo-spin associated with the electron’s degree of freedom on the sublattice, while $\hat{s}_i$ represents the Pauli matrices of the electron’s real spin. $\hat{s}_0 $ and $\hat{\sigma}_0$ are the unit $2\times2 $ identity matrix and $\hat{\sigma}_\pm=(\hat{\sigma}_0\pm\hat{\sigma}_z)/2$. The quantities $\Delta_{\beta \zeta}^{\pm} = \Delta/2 \pm \zeta \lambda_\beta$ represent the valley-dependent spin splittings in the conduction ($\beta = c$) and valence ($\beta = v$) bands, with $\Delta$ denoting the intrinsic band gap, and $\lambda_c$  ($\lambda_v$) the iSOC strengths for the conduction (valence) bands~\cite{xiao2012, kormanyos2014}. The parameter $a$ is the lattice constant, $t$ the effective hopping energy~\cite{lu2013}, and $\lambda_R = \gamma_R \Delta / (2 a t)$ quantifies the RSOC strength, with $\gamma_R$ the RSOC parameter~\cite{xu2003, yao2017}.

The four eigenvalues $E = E_{\beta s}^\zeta(\mathbf{k})$, where $s = \pm$ labels the spin-split subbands, are obtained by diagonalizing the Hamiltonian. The corresponding eigenstates are denoted by $|\mathbf{k}; \alpha\rangle$.
The four eigenvalues $E = E_{\beta b}^\zeta (\mathbf{k})$, with $ b = (1,2)$ the different spin subbands, are the solutions of the diagonalized equation of the matrix, which reads
\begin{equation}
E^4-A_2E^2+A_1E+A_0=0,
\end{equation}
with
\begin{align}
A_2=&\Delta^2/2+\lambda_c^2+\lambda_v^2+2a^2t^2k^2+4\lambda_R^2, \tag{2a}\\
A_1=&\Delta(\lambda_v^2-\lambda_c^2)+4\lambda_{R}^2(\lambda_c-\lambda_v), \tag{2b}\\
A_0=&(\Delta^2/4+a^2t^2k^2)^2-\Delta^2[\lambda_c^2+\lambda_v^2]/4\notag \\
&+\lambda_c\lambda_v[\lambda_c\lambda_v-2a^2t^2k^2]+4\lambda_{R}^2(\Delta/2+\zeta \lambda_c)\notag \\
&\times(\Delta/2+\zeta \lambda_v). \tag{2c}
\end{align}
The corresponding eigenfunctions are given by
\begin{align}
|\mathbf{k};\alpha\rangle=\mathbb{C}_{\beta b}^\zeta (\mathbf{k})[h_1,h_2,h_3,h_4]e^{i\mathbf{k \cdot r}},
\end{align}
with $\alpha=(\zeta, b, \beta)$, and
\begin{align}
h_1=& i\lambda_{R}[(\zeta+1)a^2t^2(k^-)^2-(\zeta-1)O_{c \zeta}^{-}O_{v \zeta}^{-}], \tag{3a}\\
h_2=& atk^-[a^2t^2k^2+O_{c \zeta}^{+}O_{v \zeta}^{-}], \tag{3b}\\
h_3=& -2i\lambda_{R}atk^{-}O_{c \zeta}^{+}, \tag{3c} \\
h_4=& -O_{c \zeta}^{+}O_{c \zeta}^{-}O_{v \zeta}^{-}-(\zeta+1)^2\lambda_{R}^2O_{c \zeta}^{+}-a^2t^2k^2O_{c \zeta}^{-},\tag{3d}
\end{align}
where $\mathbb{C}_{\beta b}^\zeta (\mathbf{k})= (|h_1|^2+|h_2|^2+|h_3|^2+|h_4|^2)^{-1/2}$, $O_{c \zeta}^{\pm}=\Delta_{c \zeta}^{\pm}-E$, and $O_{v \zeta}^{\pm}=\Delta_{v \zeta}^{\pm}+E$.

For ML-TMD systems without RSOC, the spin operator along the $z$ direction $\hat{P}_z$, $\hat{P}_z=\hbar \hat{\sigma}_0 \otimes\hat{s}_z/2$, commutes with the Hamiltonian, both $\hat{H}_{0}$ and ${H}_{\rm {iSOC}}$. In this case, the spin index $s = \pm 1$ is commonly used to label the spin-up and spin-down subbands, where $s$ serves as a good quantum number. 

However when RSOC is present, $\hat{P}_z$ no longer commutes with the Hamiltonian $\hat{H}_R$,
\begin{align}
[\hat{H}_{R},\hat{P}_z]=&\lambda_R([\zeta \hat{\sigma}_x \otimes \hat{s}_y,\hat{P}_z] -[\hat{\sigma}_y \otimes \hat{s}_x,\hat{P}_z]) \neq 0.
\end{align}
Using spin index $s = \pm 1$ to label subbands is not appropriate, and we use $b = (1, 2)$ instead to label the two distinct subbands. 

%=& i \lambda_R\hbar (\zeta \hat{\sigma}_x \otimes \hat{s}_x+\hat{\sigma}_y \otimes \hat{s}_y) 

At finite temperatures, the electron number
conservation law is
\begin{align}
n_e&=\sum_{\zeta,b,\mathbf{k}}f[E_{c b}^\zeta(\mathbf{k})],
\end{align}
with $n_{e}$ the electron density for electrons in an $n$-type sample and $f(x)=\{1+{\rm exp}[(x-\mu)/(k_BT)]\}^{-1}$
the Fermi-Dirac function.  $\mu$ is the chemical
potential (or equivalently at zero temperature, the Fermi energy).

Within the random phase approximation (RPA), the dynamical dielectric function for intravalley $n$-type ML-TMDs can be expressed as \cite{giuliani2008quantum, kotov2012electron}
\begin{equation}
\epsilon(\mathbf{q}, \omega) = 1 - v_q \sum_{\mathbf{k}} \sum_{\zeta b b'} F_{c, b b'}^\zeta(\mathbf{k}, \mathbf{q}) \, \Pi_{b b'}^\zeta(\mathbf{k}, \mathbf{q}, \omega). \label{epsilon}
\end{equation}
The wave vector $\mathbf{q} = (q_x, q_y)$ represents the momentum transfer in an electron-electron scattering process. $v_q = 2\pi e^2 / (\epsilon_{\rm{env}} q)$ is the 2D Fourier transform of the Coulomb potential, with $\epsilon_{\rm{env}} = (\varepsilon_{\rm{sub}}+\varepsilon_0)/2$ denoting the effective dielectric constant of the surrounding environment \cite{hwang2007dielectric} [see Sec. I in Supplementary Material (SM) \cite{Supplementarymaterial}]. Here, $\varepsilon_{\rm{sub}}$ is the dielectric constant of the substrate, and $\varepsilon_0$ is the vacuum permittivity.

The structure factor $F_{c, b b'}^\zeta(\mathbf{k}, \mathbf{q})$ accounts for the wave-function overlap and is defined as
\begin{align}
F_{c, b b'}^\zeta(\mathbf{k}, \mathbf{q}) =&|\langle\mathbf{k+\mathbf{q}},\alpha'|e^{-i\mathbf{q}\cdot\mathbf{r}}|\mathbf{k},\alpha\rangle|^2 \notag \\
=&[\mathbb{C}_{c b ^\prime}^{\zeta}(\mathbf{k+q}) \mathbb{C}_{c b}^\zeta(\mathbf{k})]^2\sum_{i=1}^4 h_i(\mathbf{k})h_i(\mathbf{k+q}) \notag \\
&\times\sum_{j=1}^4 h_{j}(\mathbf{k+q})h_{j}(\mathbf{k}).
\end{align}
The lowest-order density–density response function (polarization function) for intravalley $n$-type ML-TMDs is given by
\begin{align}
\Pi_{b b'}^\zeta(\mathbf{k}, \mathbf{q}, \omega) = 
\frac{f[E_{c b'}^\zeta(\mathbf{k} + \mathbf{q})] - f[E_{c b}^\zeta(\mathbf{k})]}
{E_{c b'}^\zeta(\mathbf{k} + \mathbf{q}) - E_{c b}^\zeta(\mathbf{k}) + \hbar \omega + i \eta}, \label{pol}
\end{align}
In the presence of RSOC, the polarization function in the $\zeta~(K/K')$ valley is given by $\sum_{ b b'}\Pi_{b b'}^{\zeta}=\Pi_{1 1}^{\zeta}+\Pi_{22}^{\zeta}+\Pi_{12}^{\zeta}+\Pi_{21}^{\zeta}$. Here, $\Pi_{1 1}^{\zeta}$ and $\Pi_{22}^{\zeta}$ correspond to intrasubband transitions, while $\Pi_{12}^{\zeta}$ and $\Pi_{21}^{\zeta}$ correspond to intersubband transitions.

The plasmon modes are obtained by solving Eq.\ \eqref{epsilon} for $\text{Re}[\epsilon(\mathbf{q}, \omega)] =$ 0, corresponding to the collective charge oscillations \cite{nozieres1958dielectric}. 
The electron energy loss function (EELF), which characterizes the energy absorbed by a fast electron traversing the material, is given by \cite{nozieres1958dielectric,egerton2011electron}
\begin{align}
    L(\mathbf{q}, \omega)&=-\text{Im}[\frac{1}{\epsilon(\mathbf{q}, \omega)}]\notag \\
    &=\frac{\text{Im}[\epsilon(\mathbf{q}, \omega) ]}{\{\text{Im}[\epsilon(\mathbf{q}, \omega) ]\}^2+\{\text{Re}[\epsilon(\mathbf{q}, \omega) ]\}^2}.
\end{align}
To obtain the real and imaginary parts of the dielectric function, we use the Dirac identity $\lim_{\eta \to 0} [1/(x \pm i\eta)] = \mathcal{P}\{1/x\} \mp i\pi \delta(x)$,  where $\mathcal{P}\{1/x\}$ denotes the principal value and $\delta(x)$ is the Dirac delta function. We use the Lorentzian broadening approximation, replacing the $\delta(x)$ function via
$\delta(x)\rightarrow (\Gamma/\pi)/(x^2+\Gamma^2)$, where broadening width $\Gamma=\hbar/\tau$ is defined by the lifetime $\tau$. While broadening width $\Gamma$ reduces intensity and sharpness of EELF, it does not shift the positions of plasmon modes \cite{yan2013damping}(see Sec. II in SM for details \cite{Supplementarymaterial}). We use a value of $\Gamma$ = 0.1 meV for numerical calculations.

\begin{figure}[t]
\centering
\includegraphics[trim=1 1 1 1, clip,width=1\linewidth]{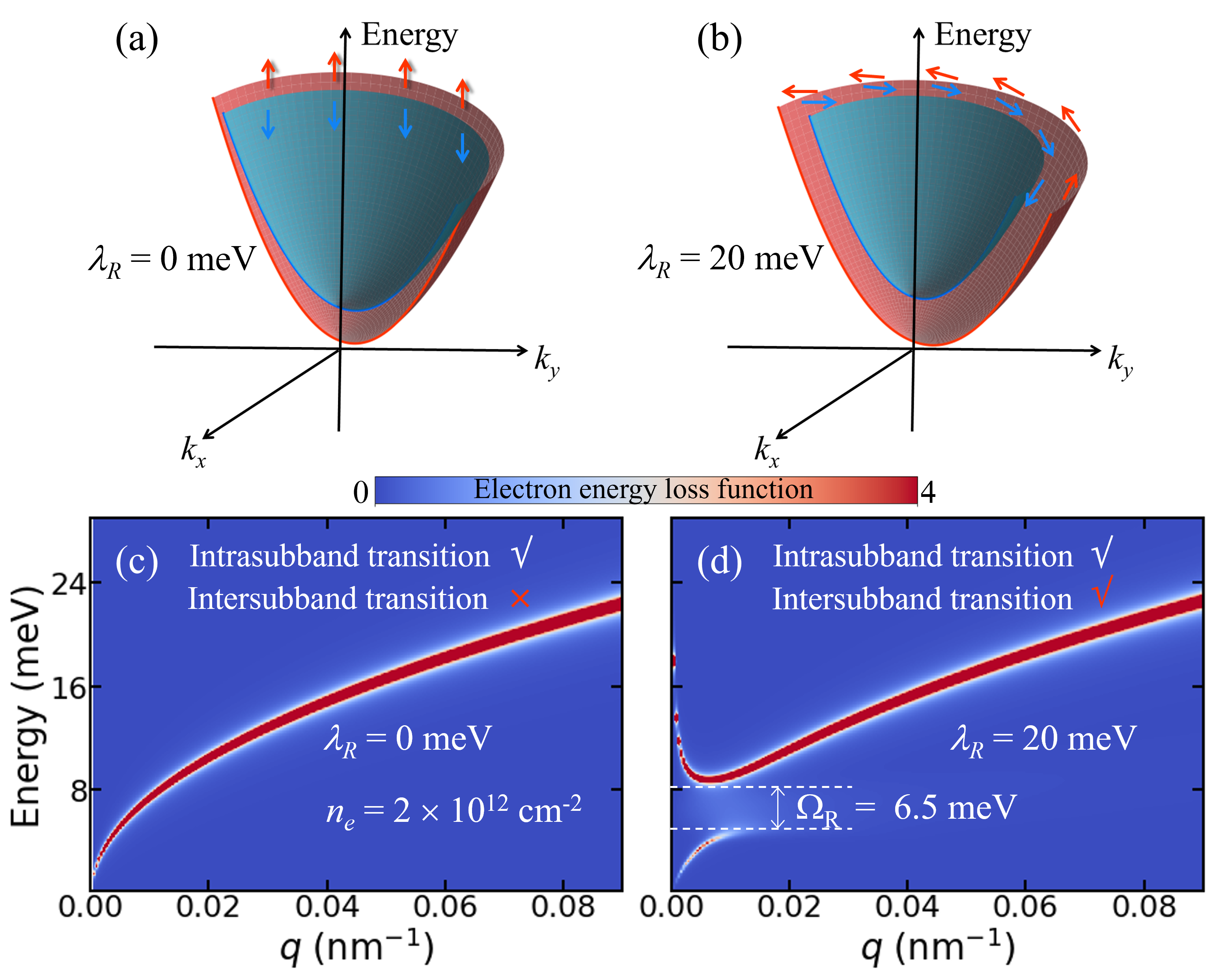}
\caption{
Schematic conduction band profiles of ML-MoS$_2$ in the $K$ valley and the corresponding electron energy loss function (EELF) at fixed electron density $n_e= 2 \times 10^{12}~\mathrm{cm}^{-2}$. (a) and (c) without RSOC; (b) and (d) including RSOC. Red and blue arrows indicate the spin textures in different subbands. A characteristic minimum energy gap $\Omega_R$ between the bottom of the optical plasmon mode and the top of the acoustic plasmon mode, serves as a fingerprint of the RSOC.
}
\label{Fig1}
\end{figure}

\begin{figure*}
\centering{}\includegraphics[width=16cm]{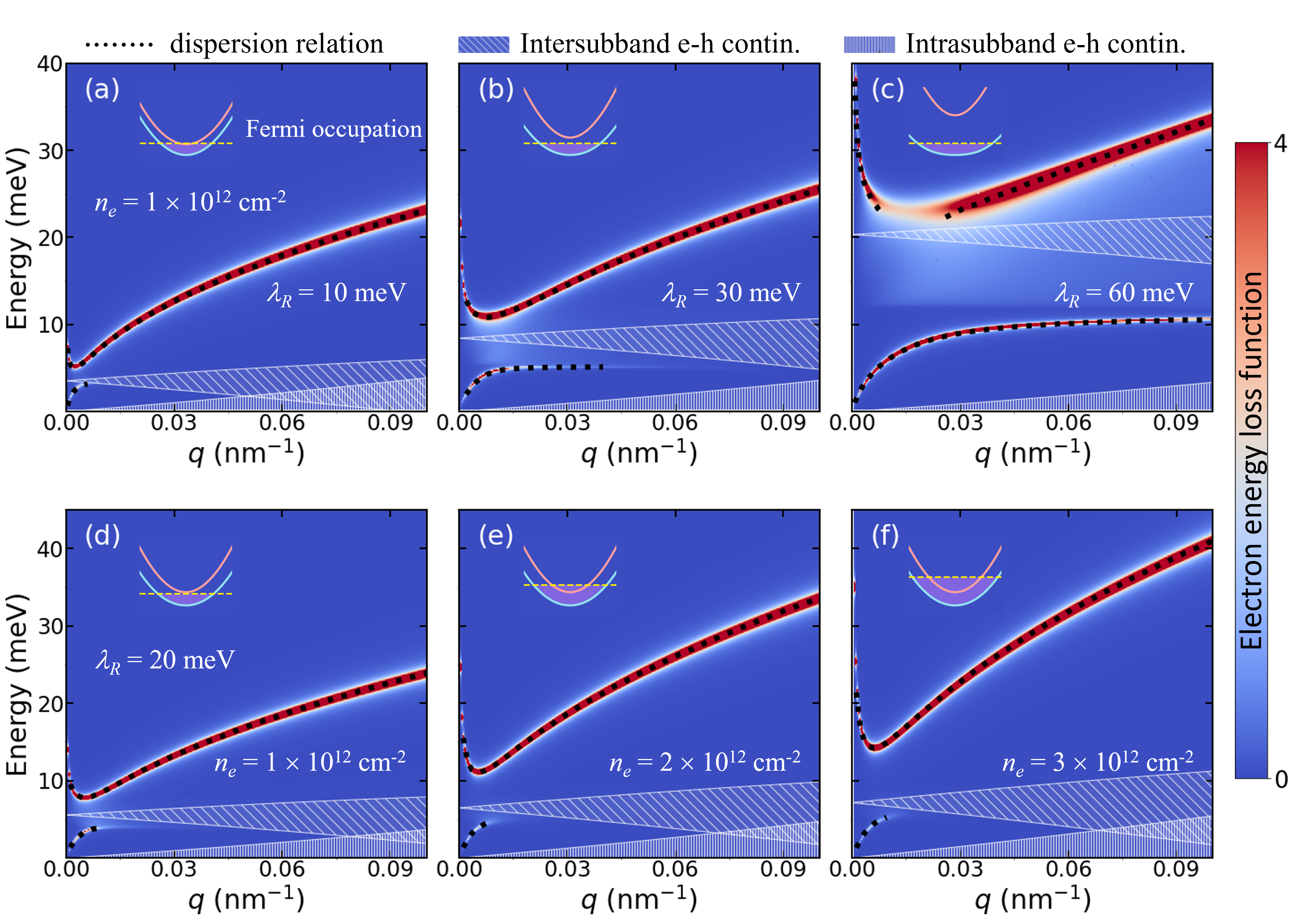}
\caption{
The EELF, plasmon dispersion, and corresponding Fermi occupation in $n$-type ML-MoS$_2$. (a)$-$(c) For fixed electron density of $n_e= 1 \times 10^{12}~\mathrm{cm}^{-2}$, with RSOC strengths $\lambda_R =$ 10, 30, and 60~meV, as labeled.
(d)$-$(f) For fixed RSOC strength $\lambda_R$ =20~meV, with electron densities $n_e= 1 \times 10^{12}~\mathrm{cm}^{-2}$, $2 \times 10^{12}~\mathrm{cm}^{-2}$, and $3 \times 10^{12}~\mathrm{cm}^{-2}$, as labeled. The white shaded regions indicate the intra- and intersubband $e$-$h$ continua.
 }\label{Fig2}
\end{figure*}

{\it Results and discussion. } 
In this Letter, we take monolayer molybdenum disulfide (ML-MoS$_2$) as an illustrative example. We consider an $n$-type ML-MoS$_2$ placed on a silicon dioxide (SiO$_2$) substrate. The RSOC strength $\lambda_R$ can be tuned via an external electrical field and is treated as an input parameter in our analysis. The following material parameters for ML-MoS$_2$ \cite{xiao2012,kormanyos2014,liu2013three} are used in our numerical calculations: $a=3.193$~\text{\AA}, $t=1.1~\text{eV}$, $\Delta=0.83$~eV, $\lambda_c=-1.5$~meV, $\lambda_v =$~74 meV. The parameters of other typical ML-TMDs~\cite{hien2020magneto}, such as ML-MoSe$_2$, ML-WS$_2$, ML-WSe$_2$, etc., can also be considered in this model. All calculations are performed at a temperature of $T = 10$~K. It should be noted that temperature would also affect the plasmon properties due to the smoothness of the Fermi-Dirac distribution and the blueshifts of the energies required for electronic transitions between spin-split subbands. Thereby, the minimum energy gap $\Omega_R$ separating the optical and acoustic plasmon modes is enhanced at $T =$ 300 K (see Fig. S2 in the SM \cite{Supplementarymaterial}).

Figure~\ref{Fig1} shows schematic conduction band profiles of ML-MoS$_2$ in the $K$ valley and the corresponding electron energy loss function $L(\mathbf{q}, \omega)$, at fixed electron density. 
The EELF characterizes the energy absorbed by a fast electron traversing the material. Cases with and without RSOC are compared.

In the absence of RSOC, as shown in Figs.~\ref{Fig1}(a) and ~\ref{Fig1}(c), the spin index is a good quantum number. The conduction subbands are characterized by distinct spin-up and spin-down states, and intersubband transitions are forbidden. Consequently, in Fig.~\ref{Fig1}(c) only a plasmon mode appears in the EELF that is induced by intrasubband transitions.

Figures \ref{Fig1}(b) and \ref{Fig1}(d) show the corresponding results in the presence of RSOC for which the spin index is no longer a good quantum number. RSOC induces not only spin splitting but also spin hybridization, resulting in subbands with opposite helical spin textures. The hybridization enables additional intersubband transitions, and this leads to the appearance of both optical and acoustic plasmon modes. The distribution of the optical and acoustic plasmon modes resembles Rabi-type splitting. This results in a finite minimum energy gap between optical and acoustic plasmon modes in the EELF at finite electron density and RSOC strength, and this can serve as a spectral signature of RSOC strength. The magnitude of the gap, denoted by $\Omega_R$, provides a direct and quantitative indicator of the RSOC strength.

To explore further the influence of RSOC strength and electron density on the plasmon modes of $n$-type ML-MoS$_2$, Figs.~\ref{Fig2}(a)$-$(c) present the EELF, plasmon dispersion and the corresponding Fermi occupation of $n$-type ML-MoS$_2$, for increasing RSOC strengths at fixed electron density $n_e= 1 \times 10^{12}~\mathrm{cm}^{-2}$. The white shaded regions indicate the intra- and intersubband electron-hole ($e$-$h$) continua within the conduction band.  These correspond to the regions where the imaginary part of the polarization function is nonzero ($-\text{Im}[\Pi_{b b'}^\zeta(\mathbf{k}, \mathbf{q}, \omega)]\neq 0$), signifying regions of Landau damping where external fields are absorbed \cite{giuliani2008quantum}. Consequently, collective excitations are heavily damped within these continua. The black dotted curves in Fig.~\ref{Fig2}, representing the plasmon dispersion of $n$-type ML-MoS$_2$ obtained by solving $\text{Re}[\epsilon(\mathbf{q}, \omega)] =$ 0, coincide with regions of high spectral weight in the EELF. These curves reveal two distinct plasmon branches, optical and acoustic plasmon modes, with the minimum energy gap between them denoted by $\Omega_R$. The energy gap $\Omega_R$ originates from RSOC-induced spin hybridization, which opens additional intersubband transition channels. As RSOC increases, the Fermi occupations show a widening energy separation between subbands, leading to a corresponding increase in $\Omega_R$. Since RSOC can be tuned by an external electric field~\cite{patel2022electric} or heterostructuring~\cite{feng2025giant}, the plasmonic properties can be effectively modulated by adjusting the RSOC strength. Thus RSOC can serve as a potent and tunable parameter for controlling plasmonic properties in 2D materials and van der Waals heterostructures. 

Figures~\ref{Fig2}(d)$-$\ref{Fig2}(f) present the EELF, plasmon dispersion, and the corresponding Fermi occupation of $n$-type ML-MoS$_2$ for increasing electron densities at a fixed RSOC strength. As electron density is increased, the minimum energy gap $\Omega_R$ increases, consistent with the reported plasmon dispersion $\omega(q)\sim q^{1/2} n_e^{1/2}$ in ML-MoS$_2$~\cite{scholz2013plasmons}. This demonstrates that electron density can serve as an additional tuning parameter for manipulating plasmon modes. Altogether, Fig.\ \ref{Fig2} shows how RSOC strength and electron density together influence the plasmon modes, and how the minimum energy gap between the two modes can be tuned. Here, the EELF obtained from our theoretical calculations can be directly accessed via experimental electron energy loss spectroscopy (EELS)~\cite{van2011effect,macmahon2015layered,dileep2016layer,forcherio2017electron}. In particular, previous studies have experimentally obtained the plasmon dispersion relation of ML-TMDs using EELS~\cite{van2011effect}. Thus, our theoretical predictions can be validated experimentally.

\begin{figure}
\centering{}\includegraphics[width=0.9\linewidth]{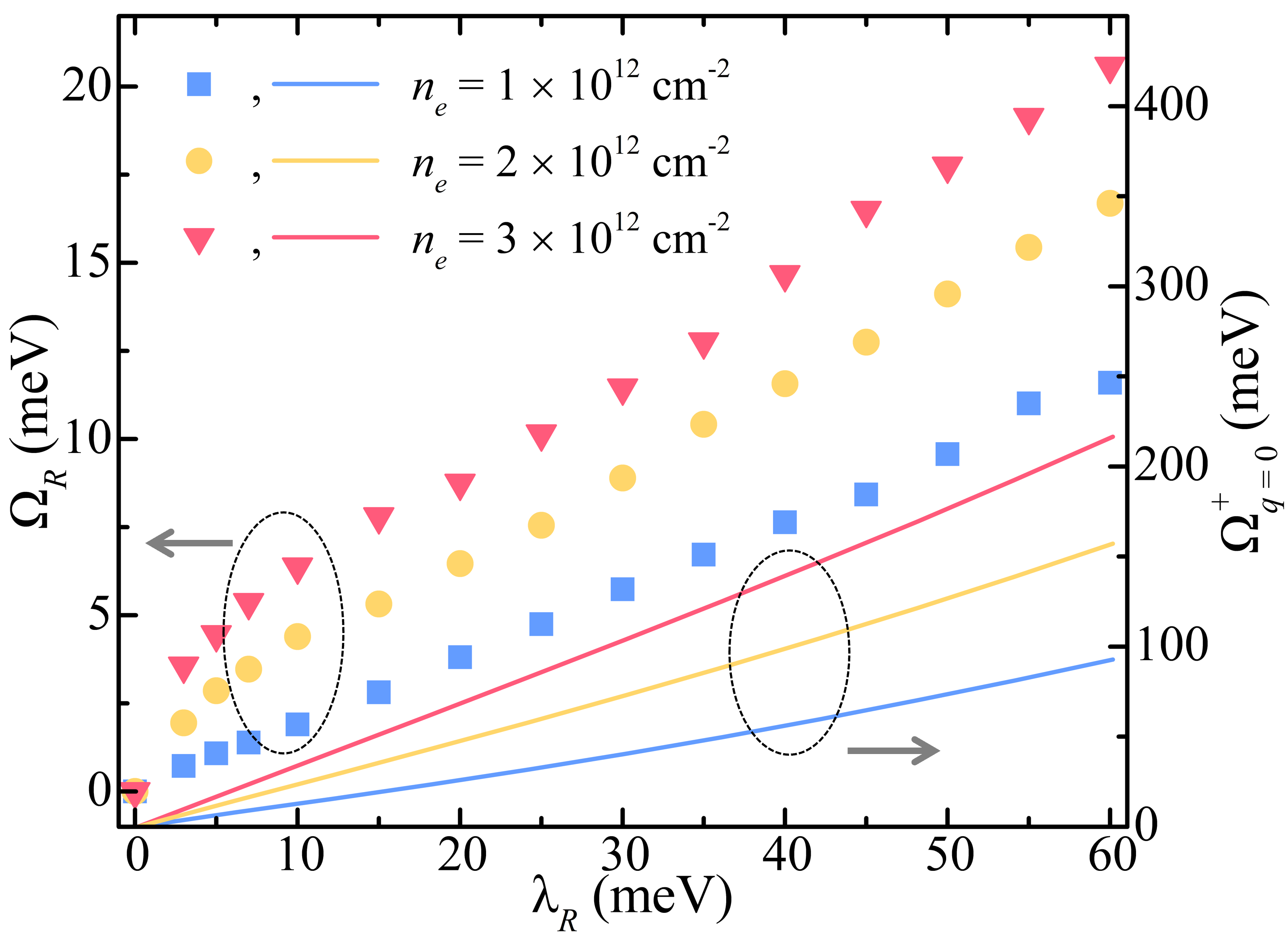}
 \caption{The minimum energy gap $\Omega_R$ separating the two plasmon modes and the plasmon energy $\Omega^+_{q=0}$ of the optical plasmon mode at $q\rightarrow0^+$, as a function of RSOC strength $\lambda_R$, for an $n$-type ML-MoS$_2$ at different electron densities $n_e =1 \times 10^{12}~\mathrm{cm}^{-2}$ (blue), $2 \times 10^{12}~\mathrm{cm}^{-2}$ (yellow), and $3 \times 10^{12}~\mathrm{cm}^{-2}$ (red).}\label{Fig3}
\end{figure}

To directly demonstrate the relationship between the RSOC strength $\lambda_R$ and the minimum energy gap $\Omega_R$ separating the optical and acoustic plasmon modes, Fig.~\ref{Fig3} shows the minimum energy gap $\Omega_R$ as a function of $\lambda_R$ for $n$-type ML-MoS$_2$, at different electron densities $n_e$. The energy gap $\Omega_R$ is extracted from the plasmon dispersion relation. One sees that at a fixed electron density $n_e$, $\Omega_R$ increases monotonically with increasing $\lambda_R$. However, a change in functional dependence is observed at small RSOC strength $\lambda_R$, due to the presence of the $e$-$h$ continuum. 
%As shown in Fig.~\ref{Fig2}(a), at a small RSOC strength $\lambda_R=10$ meV, the acoustic plasmon mode enters the e-h continuum, affecting the energy gap. 
At weak RSOC, $\lambda_R=10$~meV [see Fig.~\ref{Fig2}(a)], the acoustic plasmon mode enters the $e$-$h$ continuum before reaching the gap minimum. In such a case, we take $\Omega_R$ to be the minimum energy gap between the boundary of the $e$-$h$ continuum and the optical plasmon. This changes the functional dependence of $\Omega_R$ on $\lambda_R$.
For the electron density $n_e =1 \times 10^{12}~\mathrm{cm}^{-2}$, a change in the functional dependence also appears at a large RSOC strength, this time caused by the entry of the optical plasmon mode into the $e$-$h$ continuum, as shown in Fig.~\ref{Fig2}(c). Figure~\ref{Fig3} also gives
the plasmon energy of the optical plasmon mode (upper plasmon mode) $\Omega^+_{q=0} $ at $q\rightarrow0^+$ as a function of RSOC strength $\lambda_R$. For a fixed electron density $n_e$, $\Omega^+_{q=0} $ has a roughly linear relationship with $\lambda_R$ which is attributed to the increase of the energy required for electronic transitions between spin-split subbands for increasing $\lambda_R$. Furthermore, at a fixed RSOC strength $\lambda_R$, a large electron density also results in a higher $\Omega^+_{q=0}$. Thus, $\lambda_R$ can strongly affect the values of $\Omega_R$ and $\Omega^+_{q=0}$. Usually, the optical plasmon mode $\Omega^+_{q=0}$ can be easily measured experimentally. However, the spectral strength in the EELF is not obvious. For a fixed RSOC strength $\lambda_R$, 
Fig.~\ref{Fig3} shows that increasing the electron density results in a larger $\Omega_R$. 
These findings demonstrate that, for a fixed electron density, the RSOC strength $\lambda_R$ can be determined from the minimum energy gap $\Omega_R$ measured by EELS. 

In addition, since the RSOC strength in ML-TMDs can be obtained from ARPES measurements~\cite{zribi2022unidirectional} and the electron density can be precisely controlled via gate voltage, the corresponding minimum energy gap $\Omega_R$ could be determined using Fig.~\ref{Fig3}. In this way, an experimentally determination of  $\Omega_R$ would allow direct independent experimental verification of our results.

{\it Conclusion.} In this Letter, we have presented a theoretical analysis of plasmonic properties of $n$-type ML-TMDs in the presence of Rashba spin-orbit coupling (RSOC). We have proposed a plasmonic method to accurately probe the Rashba effect in ML-TMDs by analyzing features in the electron energy loss functions and dispersion relations. Specifically, we identify a spectral minimum energy gap $\Omega_R$ separating the two plasmon modes to serve as a direct spectral signature of inter-subband transitions enabled by RSOC. This feature has been shown to directly depend on the RSOC strength and electron density, and is accessible in electron energy loss spectroscopy (EELS) measurements. Our results reveal that the strength and visibility of $\Omega_R$ are governed by how RSOC modifies spin-dependent band structures and alters the dielectric response of the system. These findings establish EELS as a noninvasive, tunable, and precise technique for determining the RSOC, providing a practical approach for large-scale characterization and control of spin-orbit phenomena in 2D materials and heterostructures, with promising implications for spintronic and plasmonic technology platforms of tomorrow. In addition, the present conclusions are not limited to ML-TMDs but remain valid also for other 2D materials. In principle, as long as the material possesses RSOC, similar physical results can be expected \cite{Andreas2012}.

{\it Acknowledgments.} We thank H. M. Dong for important discussions. 
This work was supported by the National Natural Science Foundation of China (NSFC) (Grants No. 12364009, No. U2230122, and No. U2067207), Yunnan Fundamental Research Projects (Grant No. 202301AT070120), and the Scientific Research Fund Project of Yunnan Education Department (Grant No. 2025Y0066), the EU-funded DYNASTY project (Grant No. 101079179) and
 the Research Foundation-Flanders (FWO-Vlaanderen). 
 Z.H.T and Y.L. acknowledge China Scholarship Council support. Y.M.X. was supported through the Xingdian Talent Plans for Young Talents of Yunnan Province (Grant No. XDYC-QNRC-2022-0492).

{\it Data availability - } The data that support the findings of this article are not publicly available. The data are available from the authors upon reasonable request.

\bibliography{ref}
\end{document}